\documentclass[12pt,draftcls,a4paper,onecolumn]{IEEEtran}

\usepackage[latin1]{inputenc}
\usepackage{graphicx}
\usepackage{amssymb}
\usepackage{amsmath}

\begin{document}

\title{Exact Failure Frequency Calculations for Extended Systems}

\author{Annie~Druault-Vicard~and~Christian~Tanguy%
\thanks{A. Druault-Vicard and C. Tanguy are with France Telecom Division R\&D CORE/MCN/OTT,
38--40 rue du G\'{e}n\'{e}ral Leclerc, 92794 Issy-les-Moulineaux Cedex 9, France.}%
\thanks{E-mail: annie.vicard@orange-ftgroup.com, christian.tanguy@orange-ftgroup.com.}}

\maketitle

\begin{abstract}
This paper shows how the steady-state availability and failure frequency can be calculated in a single pass
for very large systems, when the availability is expressed as a product of matrices. We apply the general
procedure to $k$-out-of-$n$:G and linear consecutive $k$-out-of-$n$:F systems, and to a simple ladder network
in which each edge and node may fail. We also give the associated generating functions when the components
have identical availabilities and failure rates. For large systems, the failure rate of the whole system is
asymptotically proportional to its size.

This paves the way to ready-to-use formulae for various architectures, as well as proof that the differential
operator approach to failure frequency calculations is very useful and straightforward.
\end{abstract}


%\markboth{}{A. DRUAULT-VICARD and C. TANGUY}
\markboth{}{}

\section*{Acronyms\footnote{The singular and plural of an acronym are always spelled the same.}}

\begin{tabular}{ll}
GVI & grouped variable inversion (method) \\
SVI & single variable inversion (method) \\
IE & inclusion-exclusion (principle) \\
OBDD & Ordered Binary Decision Diagram
\end{tabular}

\section*{Notation}

\begin{tabular}{ll}
$p_i, q_i$ & [success, failure] probability of component $i$ \\
& ($q_i = 1 - p_i$)  \\
$p, q$ & implies $p_i =p$, $q_i = q$ (for edges). \\
$\rho$ & identical availability of nodes (when $\rho \neq p$). \\
$\lambda_i$, $\mu_i$ & [failure, repair] rate of component $i$ \\
$\lambda$, $\mu$ & common [failure, repair] rate of components \\
$A$ & steady-state availability of the system \\
$U$ & steady-state unavailability of the system \\
$\overline{\nu}$ & mean failure frequency of the system \\
$\overline{\lambda}$ & mean failure rate of the system ($\overline{\nu} =  A \, \overline{\lambda}$)\\
%${\mathbf 1}_k$ & identity matrix in a space of dimension $k$ \\
%${\rm Det}(M)$ & determinant of matrix $M$ \\
$M'$ & $\displaystyle \left( \sum_i \, \lambda_i \, p_i \,
\frac{\partial}{\partial p_i} \right) M$ \\
${\mathcal G}(z)$ & generating function for the availability \\
$\widehat{{\mathcal G}}(z)$ & generating function for the failure frequency \\
$A_{k,n}$ & availability of a $k$-out-of-$n$:G system
\end{tabular}

\section{Introduction}

\PARstart{S}{teady-state} system availability and failure frequency are important performance indices of a
repairable system \cite{Shooman68,Singh77,Colbourn,KuoZuo}, from which other key parameters such as the mean
time between failures, average failure rate, Birnbaum importance, etc. may be deduced. In the steady-state
regime, the frequency of system failure was first calculated by a cut-set \cite{Singh74} or a tie-set
approach \cite{Singh75} in the case of statistically independent failures, which will also be considered
here. These approaches are based on the inclusion-exclusion (IE) principle, where the failure or repair rates
(more generally, the inverses of the mean down or up times), are adequately given for each term of the
relevant expansion.

When all the terms of its IE expansion are kept, the exact availability is obtained as a function of each
component availability. Several papers have provided a few simple recipes, describing how the system failure
frequency and the failure rate can then be derived \cite{Schneeweiss81,Shi81,Yuan86}. Recent refinements have
been proposed when availability expressions are obtained from various instances (SVI, GVI) of
sum-of-disjoint-products algorithms \cite{Amari00}. All these formal calculations boil down to a simple fact:
the failure frequency may be derived from the availability through the application of a linear differential
operator \cite{Schneeweiss83,Hayashi91}. This requires knowledge of the exact availability, which is hard to
come by except for trivially small networks, and may have hindered the use of this method.

Unsurprisingly, several algorithms have been put forward, in which availability and failure frequency are
computed side by side in a common procedure: triangle-star transformation \cite{Gadani}, OBDD calculations
\cite{Chang04}, and another instance of differential operator calculations \cite{Hayashi06}.

In this paper, we want to promote the differential operator method for the calculation of the failure
frequency by showing it gives the exact result for numerous, widely used configurations, with an {\em
arbitrary large} number of components. We take advantage of recent results establishing that the availability
of recursive networks may be expressed as a product of transfer matrices that take each edge and node
availabilities exactly into account \cite{Tanguy06a,Tanguy06b,Tanguy06c}.

Our paper is organized as follows. In Section \ref{General procedure}, we show how the failure frequency of a
system may generally be deduced from the steady-state availability when the latter is expressed by a product
of transfer matrices. We first apply this method in Section \ref{KoutofN}, which is devoted to $k$-out-of-$n$
systems (either $k$-out-of-$n$:G or linear consecutive $k$-out-of-$n$:F ones) with distinct components.
Section \ref{SimpleLadder} provides a generic example for the two-terminal failure frequency of a simple
ladder network, which has been solved recently for arbitrary edge and node availabilities \cite{Tanguy06a};
the same procedure could easily be used for more complex networks and their all-terminal reliability too
\cite{Tanguy06b,Tanguy06c}. In each configuration, we pay attention to the case of identical components, for
which the common availability is $p$ (for edges) and $\rho$ (for nodes). For large systems, we show that the
asymptotic failure rate has a linear dependence with size, and is given by derivatives of the largest
eigenvalue of the unique transfer matrix with respect to $p$ and $\rho$. We conclude by a brief outlook.

\section{General procedure}
\label{General procedure}

In many systems, as will be explicitly shown in the following sections, the availability $A$ (or the
unavailability $U$) is given by an expression of the form
\begin{equation}
A =  {\mathbf v}_L \, M_n \, \cdots \, M_1 \, {\mathbf v}_R \, ,
\end{equation}
where $M_k \ (1 \leq k \leq n)$ is a transfer matrix, the elements of which are multilinear polynomials of
individual component availabilities, and where ${\mathbf v}_L$ and ${\mathbf v}_R$ are two vectors in which
these availabilities do not appear. The mean failure frequency $\overline{\nu}$ is obtained from
\cite{Schneeweiss83,Hayashi91}
\begin{equation}
\overline{\nu} =  \sum_i \lambda_i \, p_i \, \frac{\partial A}{\partial p_i} = \sum_i \mu_i \, q_i \, \,
\frac{\partial U}{\partial q_i}. \label{nuGenerale}
\end{equation}
In order to avoid unnecessarily heavy notation, we call $M'_k$ the matrix obtained by applying the linear
differential operator $\displaystyle \sum_i \, \lambda_i \, p_i \, \frac{\partial}{\partial p_i}$ to $M_k$.
Therefore,
\begin{eqnarray}
\overline{\nu} & = & {\mathbf v}_L \, M'_n \, M_{n-1} \,  \cdots \, M_1 \,
{\mathbf v}_R \nonumber \\
& & + {\mathbf v}_L \, M_n \, M'_{n-1} \,  \cdots \, M_1 \,
{\mathbf v}_R \nonumber \\
& & + \cdots \nonumber \\
& & + {\mathbf v}_L \, M_n \, M_{n-1} \,  \cdots \, M'_1 \, {\mathbf v}_R  \, .
\end{eqnarray}
Since $M_k$'s elements are at most linear functions of each $p_i$, the derivation of $M'_k$ is
straightforward. For instance, a matrix element $p_1 + p_2 \, p_3 - p_1 \, p_2 \, p_3$ in $M_k$ would give
rise to $\lambda_1 \, p_1 + (\lambda_2 + \lambda_3) \, p_2 \, p_3 - (\lambda_1 + \lambda_2 + \lambda_3) \,
p_1 \, p_2 \, p_3$; the recipes given in \cite{Schneeweiss81,Shi81,Schneeweiss83} fully apply.

Both availability and failure frequency may be obtained in a single pass in the following way. Let us
initialize the procedure by setting
\begin{eqnarray}
{\mathcal A}_1 & = & M_1 \, {\mathbf v}_R \, , \\
{\mathcal V}_1 & = & M'_1 \, {\mathbf v}_R \, .
\end{eqnarray}
The recursion equations are
\begin{eqnarray}
{\mathcal A}_k & = & M_k \, {\mathcal A}_{k-1} \, ,\\
{\mathcal V}_k & = & M_k \, {\mathcal V}_{k-1} + M'_k \, {\mathcal
A}_{k-1} \, ,
\end{eqnarray}
from which we deduce the final results
\begin{eqnarray}
A & = & {\mathbf v}_L \, {\mathcal A}_{n} \, ,\\
\overline{\nu} & = & {\mathbf v}_L \, {\mathcal V}_{n} \, .
\end{eqnarray}
We can now turn to a few `real-life' applications.

\section{$k$-out-of-$n$ systems}
\label{KoutofN}

$k$-out-of-$n$ systems are widely used, in various configurations; they have therefore contributed to a huge
body of literature (see \cite{KuoZuo,Chao95,Kuo94} and references therein). We start our discussion with
these systems because each transfer matrix actually refers to a single equipment only.

\subsection{$k$-out-of-$n$:G systems}
\label{KoutofN1} We first consider the simple $k$-out-of-$n$:G system, where each component has an
availability $p_i$ ($1 \leq i \leq n$). To operate as a whole, the system needs at least $k$ elements to
function. Its availability $A_{k,n}$ may be written as (see \cite{KuoZuo}, p. 244)
\begin{equation}
A_{k,n} = 1 - (1, 0, \, \cdots \, , 0)_k  \;  {\mathbf \Lambda}_n \, {\mathbf \Lambda}_{n-1} \, {\mathbf
\Lambda}_1 \; \left(
\begin{array}{c}
1 \\
1 \\
\vdots \\
1
\end{array}
\right)_{\! \! \! k} \, ,
\label{AvailabilityKout-of-N}
\end{equation}
with
\begin{equation}
{\mathbf \Lambda}_i = \left(
\begin{array}{ccccc}
q_i & p_i & 0 & \cdots & 0 \\
0 & q_i & p_i & 0 & 0\\
0 & 0 & \ddots & \ddots & 0 \\
\vdots & \vdots & 0 & q_i & p_i \\
0 & 0 & \cdots & 0 & q_i \\
\end{array}
\right)_{\! \! \! k \times k} \, . \label{transferKout-of-N}
\end{equation}
We have reduced the size of the matrix to a $k \times k$ one, instead of the original $(k+1) \times (k+1)$,
because of the nature of ${\mathbf v}_L = (1,0, \ldots, 0)_k$ and ${\mathbf v}_R$ in
eq.~(\ref{AvailabilityKout-of-N}).

The `derivative' of ${\mathbf \Lambda}_i $ is
\begin{equation}
{\mathbf \Lambda}'_i = \left(
\begin{array}{ccccc}
- \lambda_i \, p_i & \lambda_i \, p_i & 0 & \cdots & 0 \\
0 & - \lambda_i \, p_i & \lambda_i \, p_i & 0 & 0 \\
0 & 0 & \ddots & \ddots & 0 \\
\vdots & \vdots & 0 & - \lambda_i \, p_i & \lambda_i \, p_i \\
0 & 0 & \cdots & 0 & - \lambda_i \, p_i \\
\end{array}
\right)_{\! \! \! k \times k} \, ,
\end{equation}
so that the computation of the failure frequency following the method given in section \ref{General
procedure} is straightforward (care should of course be taken of the minus sign in
eq.~(\ref{AvailabilityKout-of-N})).

Let us revisit Example 7.2 of \cite{KuoZuo} (see p. 245) for the 5-out-of-8:G system with $p_i$ = 0.90, 0.89,
..., 0.83. Assuming a unique repair rate for all components, namely $\mu$, the failure rates $\lambda_i$ are
such that $\lambda_i \, p_i = \mu \, (1-p_i)$. From the procedure detailed in Section \ref{General
procedure}, we deduce $A_{5,8} = \frac{615925280183}{625000000000} \approx 0.98548045$ and a failure
frequency $\overline{\nu}_{5,8} = \frac{8012914359}{156250000000} \, \mu \approx 0.051283 \, \mu$. The
failure rate $\overline{\lambda}_{5,8} = \overline{\nu}_{5,8}/A_{5,8}$ is then equal to $0.0520382 \, \mu$.

When all components are identical ($p_i \equiv p$ and $\lambda_i \equiv \lambda$), only one transfer matrix
appears. Admittedly, $A_{k,n}$ is so simple that a matrix formulation is hardly necessary. Nonetheless, we
can give a compact expression for the generating function $\displaystyle {\mathcal G}_k(z) =
\sum_{n=0}^{\infty} \, A_{k,n} \, z^n$ (the derivation is given in the appendix):
\begin{equation}
{\mathcal G}_k(\mbox{$k$-out-of-$n$:G};z) = \frac{p^k \, z^k}{(1 - z) \, (1 - (1-p) \, z)^k} \, .
\label{GKoutofN:G}
\end{equation}
Since the generating function is a formal power-series expansion, we can apply the linear differential
operator $\lambda \, p \, \frac{\partial}{\partial p}$ to eq.~(\ref{GKoutofN:G}) so that $\displaystyle
\widehat{{\mathcal G}}_k(z) = \sum_{n=0}^{\infty} \overline{\nu}_{k,n} \, z^n$ is easily found to be
\begin{equation}
\widehat{{\mathcal G}}(\mbox{$k$-out-of-$n$:G};z) = \frac{\lambda \, k \, p^k \, z^k}{(1 - (1-p) \, z)^{k+1}}
\, , \label{GchapeauKoutofN:G}
\end{equation}
which is another formulation of the well-known result $\overline{\nu}_{k,n} = \lambda \, k \, \left( \! \! \!
\begin{array}{c}
n \\ l
\end{array}
\! \! \! \right) \, p^k \, (1-p)^{n-k}$ (eq.~(7.10) of \cite{KuoZuo}, p. 234).

\subsection{Linear consecutive $k$-out-of-$n$:F systems}
\label{KoutofN2}

These systems have been studied in many papers \cite{Chao95,Kuo94} and a recent textbook \cite{KuoZuo}. The
reliability $\widetilde{A}_{k,n}$ --- the probability of operation of a system of $n$ components, which fails
if at least $k$ {\em consecutive} elements fail --- of such a system is given by (see also eq.~(9.48) of
\cite{KuoZuo}, p. 344)
\begin{equation}
\widetilde{A}_{k,n} = (1,0, \ldots, 0)_{k} \; \widetilde{{\mathbf \Lambda}}_n \, \widetilde{{\mathbf
\Lambda}}_{n-1} \, \widetilde{{\mathbf \Lambda}}_1 \; \left(
\begin{array}{c}
1 \\
1 \\
\vdots \\
1
\end{array}
\right)_{\! \! \! k} \, ,
\end{equation}
with
\begin{equation}
\widetilde{{\mathbf \Lambda}}_i = \left(
\begin{array}{ccccc}
p_i & q_i & 0 & \cdots & 0 \\
p_i & 0 & q_i & 0 & \vdots \\
\vdots & \vdots & 0 & \ddots & 0\\
p_i & 0 & \cdots & 0 & q_i \\
p_i & 0 & 0 & \cdots & 0
\end{array}
\right)_{\! \! \! k \times k} \, .\label{transferConsecutiveKout-of-N2}
\end{equation}
Here again, we have reduced the size of the matrix and the vectors with respect to their original
formulation. Consequently,
\begin{equation}
\widetilde{{\mathbf \Lambda}}'_i = \left(
\begin{array}{ccccc}
\lambda_i \, p_i & - \lambda_i \, p_i & 0 & \cdots & 0 \\
\lambda_i \, p_i & 0 & - \lambda_i \, p_i & 0 & \vdots \\
\vdots & \vdots & 0 & \ddots & 0\\
\lambda_i \, p_i & 0 & \cdots & 0 & - \lambda_i \, p_i \\
\lambda_i \, p_i & 0 & 0 & \cdots & 0
\end{array}
\right)_{\! \! \! k \times k} \, ,
\label{transferConsecutiveKout-of-N}
\end{equation}
leading once again to a straightforward calculation of the failure frequency.

A numerical application may be found in the Lin/Con/4/11:F model, as in Example 9.6 of \cite{KuoZuo}, where
the $p_i$'s range from 0.7 to 0.9 by steps of 0.02. Assuming again that the repair rate for each equipment is
$\mu$, we get $A_{4,11} = \frac{30105385968617}{30517578125000} \approx 0.98649329$ and a failure frequency $
\overline{\nu}_{4,11} = \frac{155495836041}{30517578125000} \, \mu \approx 0.050953 \, \mu$. The
corresponding failure rate is then equal to $0.0516505 \, \mu$.

For the sake of completeness, we give the generating function for identical components is (see eq.~(2.2) of
\cite{Canfield92,Note1})
\begin{equation}
{\mathcal G}_k(\mbox{Lin/Con/$k$/$n$:F};z) = \frac{1 - (1-p)^k \, z^k}{1 - z + p \, (1-p)^k \, z^{k+1}} \, .
\label{GLinConKoutofN:F}
\end{equation}
Use of eq.~(\ref{GLinConKoutofN:F}) to obtain the failure frequency generating function is straightforward,
and will not be repeated here.

\section{Simple ladder}
\label{SimpleLadder}

\begin{figure}[thb]
\centering
\includegraphics[bb=95 275 545 565,scale=0.55,clip=TRUE]{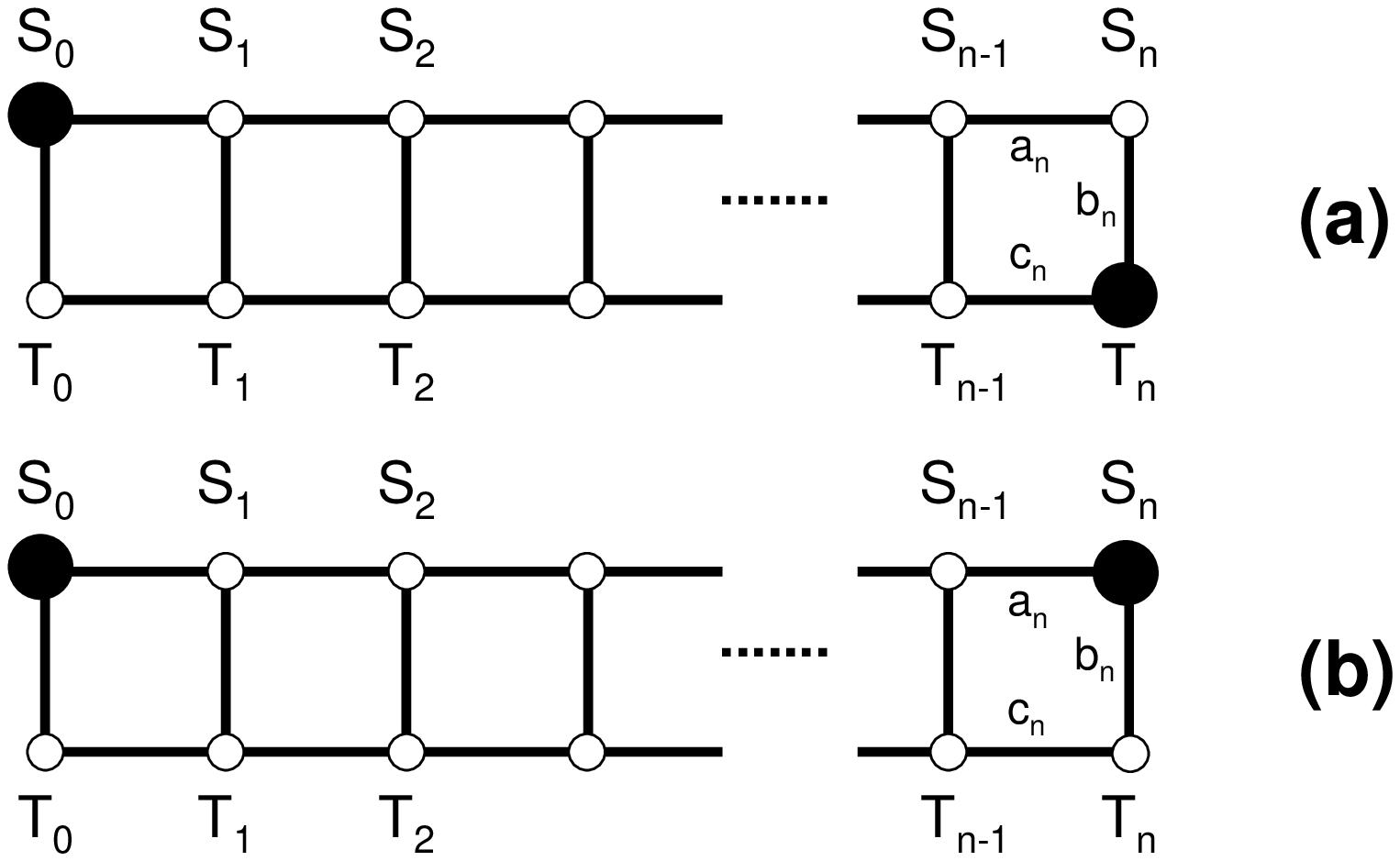}
\caption{Different source-terminal connections for a simple ladder.
Edges and nodes are indexed by their availabilities: $(a_i,b_i,c_i)$
and $(S_i,T_i)$.} \label{DeuxEchelles}
\end{figure}

We consider in this section the two-terminal availability of a simple ladder network, displayed in
Fig.~\ref{DeuxEchelles}, where successive nodes are labelled $S_i$ or $T_j$, and where the larger black dots
mark the source $s$ and terminal $t$. This network is a simplified description of a standard architecture for
long-hail communication networks: it consists in primary and backup paths, plus additional connections
between transit nodes enabling the so-called ``local protection'' policy by bypassing faulty intermediate
nodes or edges. Such an architecture of ``absolutely reliable nodes and unreliable edges,'' with up to 25
edges, was chosen as Example 5 in \cite{Heidtmann89} for a comparison of different ``sum of disjoint
products'' minimizing algorithms, or by Rauzy \cite{Rauzy03} as well as Kuo and collaborators in OBDD test
calculations \cite{Kuo99,Yeh02,Yeh02conf}. We showed \cite{Tanguy06a} that the two-terminal availability has
a beautiful algebraic structure \cite{Shier}, since its exact expression is given by a product of $3 \times
3$ transfer matrices (see eqs.~(\ref{reliabilitySn}--\ref{reliabilityMn}) below). Consequently, it can also
be determined for a network of arbitrary size.

Using the notation ${\mathcal R}_{S_n}$ (resp. ${\mathcal R}_{T_n}$) for the two-terminal availability
between $S_0$ and $S_n$ (resp. $T_n$), we find that \cite{Tanguy06a,Tanguy06c}
\begin{eqnarray}
{\mathcal R}_{S_n} & = & (1 \hskip2mm 0 \hskip2mm 0) \, M_n \,
\cdots \, M_{0} \, \left(
\begin{array}{ccc}
1 \\
0 \\
0
\end{array}
\right) \, ,
\label{reliabilitySn}\\[4mm]
{\mathcal R}_{T_n} & = & (0 \hskip2mm 1 \hskip2mm 0) \, M_n \, \cdots \, M_{0} \, \left(
\begin{array}{ccc}
1 \\
0 \\
0
\end{array}
\right) \, .
\label{reliabilityTn}
\end{eqnarray}
The transfer matrix $M_n$ is given by
\begin{equation}
M_n = \left(
\begin{array}{lll}
a_n \, S_n & b_n \, c_n \, S_n \, T_n & a_n \, b_n \, c_n \, S_n \, T_n \\
a_n \, b_n \, S_n \, T_n & c_n \, T_n & a_n \, b_n \, c_n \, S_n \, T_n \\
- a_n \, b_n \, S_n \, T_n & - b_n \, c_n \, S_n \, T_n & a_n \, (1
- 2 \, b_n) \, c_n \, S_n \, T_n
\end{array}
\right) \, .
\label{reliabilityMn}
\end{equation}
For $n=0$, we must set $a_0 = 1$, and may choose $c_0 = 0$ because it does not change the final result. It is
worth noting that all five availabilities of the $n^{\rm th}$ ``cell'' or building block of the network
appear in a single transfer matrix $M_n$, which is not sparse, contrary to the matrices of Section
\ref{KoutofN}.

Equations~(\ref{reliabilitySn}--\ref{reliabilityMn}) apply to the most general ladder in terms of individual
availabilities. If an edge or a node is missing, its reliability $p_i$ should be set to zero, and its failure
rate may be considered arbitrary, because it will not alter the final result. Similarly, if a given edge or
node is perfect, its reliability should be equal to one; its failure rate $\lambda_i$ should then, of course,
be set to zero.

The associated matrix $M'_n$ is
\begin{equation}
M'_n = \left(
\begin{array}{lll}
m'_{1 1} & m'_{1 2} & m'_{1 3} \\
m'_{2 1} & m'_{2 2} & m'_{2 3} \\
m'_{3 1} & m'_{3 2} & m'_{3 3}
\end{array}
\right) \, ,
\end{equation}
with
\begin{eqnarray}
m'_{1 1} & = & (\lambda_{a_n} + \lambda_{S_n}) \, a_n \, S_n \, , \nonumber \\
m'_{1 2} & = & (\lambda_{b_n} + \lambda_{c_n} + \lambda_{S_n} + \lambda_{T_n}) \, b_n \, c_n \, S_n \, T_n \, , \nonumber \\
m'_{1 3} & = & (\lambda_{a_n} + \lambda_{b_n} + \lambda_{c_n} + \lambda_{S_n}
+ \lambda_{T_n}) \, a_n \, b_n \, c_n \, S_n \, T_n \, , \nonumber \\
m'_{2 1} & = & (\lambda_{a_n} + \lambda_{b_n} + \lambda_{S_n} +
\lambda_{T_n}) \, a_n \, b_n \, S_n \, T_n \, , \nonumber \\
m'_{2 2} & = & (\lambda_{c_n} + \lambda_{T_n}) \, c_n \, T_n \, , \nonumber \\
m'_{2 3} & = & m'_{1 3} \, , \nonumber \\
m'_{3 1} & = & -m'_{2 1} \, , \nonumber \\
m'_{3 2} & = & -m'_{1 2} \, , \nonumber \\
m'_{3 3} & = & (\lambda_{a_n} + \lambda_{c_n} + \lambda_{S_n} + \lambda_{T_n}) \, a_n \, c_n \, S_n \, T_n -
2 \, m'_{1 3} \, . \nonumber
\end{eqnarray}

\begin{figure}[htb]
\centering
\includegraphics[scale=0.85]{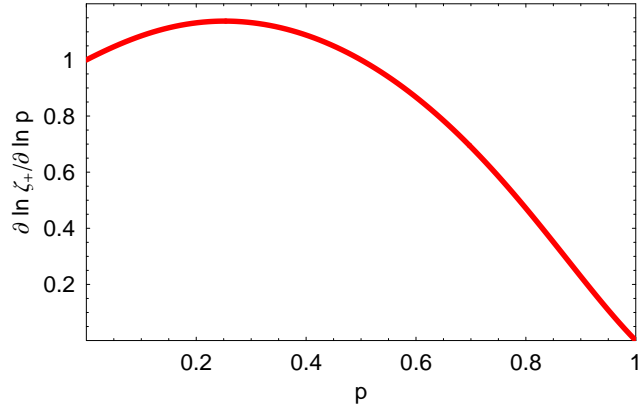}
\caption{Variation of $\partial \ln \zeta_+/\partial \ln p$ with $p$ for a simple ladder with perfect nodes
(eq.~(\ref{DLogZeta})). The maximum, reached for $p \approx 0.251641$, is about $1.13827$.}
\label{LambdaBarre3}
\end{figure}

\begin{figure}[htb]
\centering
\includegraphics[scale=0.85]{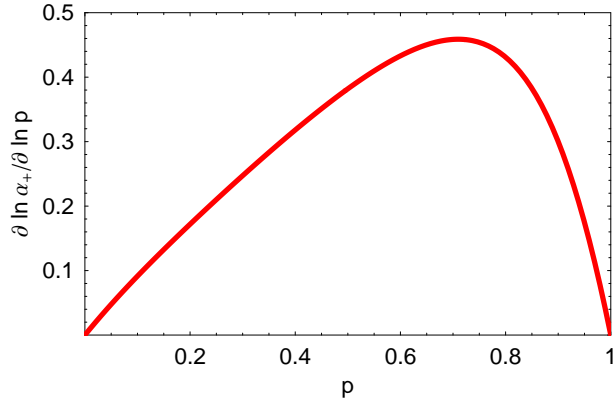}
\caption{Variation of $\partial \ln \alpha_+/\partial \ln p$ with $p$ for a simple ladder with perfect nodes
(eq.~(\ref{DLogAlpha})). The maximum, reached for $p \approx 0.709902$, is about $0.458825$.}
\label{LambdaTilde}
\end{figure}

When $a_i = b_i= c_i \equiv p$ and $S_i = T_i \equiv \rho$ the three eigenvalues $\zeta_0$ and $\zeta_{\pm}$
of the transfer matrix  are \cite{Tanguy06a}
\begin{eqnarray}
\zeta_0 & = & p \, \rho \, (1 - p \, \rho) , \label{RacinespourPetRhoX0}\\
\zeta_{\pm} & = & \frac{p \, \rho}{2} \, \left(1 + 2 \, p \,  (1 - p) \, \rho \pm \sqrt{{\mathcal B}} \right)
\, , \label{RacinespourPetRhoXpm}
\end{eqnarray}
with ${\mathcal B} = 1 + 4 \, p^2 \, \rho - 8 \, p^3 \, \rho^2 +4 \, p^4 \, \rho^2$. The two-terminal
availabilities are \cite{Tanguy06a}
\begin{eqnarray}
{\mathcal R}_{T_n}(p,\rho) & = & \frac{1}{2 \, p} \, \left(- \zeta_0^{n+1} + p \, \rho \, (1 + p \, \rho) \,
\frac{\zeta_+^{n+1} - \zeta_-^{n+1}}{\zeta_+ - \zeta_-} \right. \nonumber \\
& & \left. - (1 - 2 \, p + p \, \rho) \, p^3 \, \rho^3 \,
\frac{\zeta_+^{n} - \zeta_-^{n}}{\zeta_+ - \zeta_-} \right) \, ,\\
{\mathcal R}_{S_n}(p,\rho) & = & \frac{1}{2 \, p} \, \left(+ \zeta_0^{n+1} + p \, \rho \, (1 + p \, \rho) \,
\frac{\zeta_+^{n+1} - \zeta_-^{n+1}}{\zeta_+ - \zeta_-} \right. \nonumber \\
& & \left. - (1 - 2 \, p + p \, \rho) \, p^3 \, \rho^3 \, \frac{\zeta_+^{n} - \zeta_-^{n}}{\zeta_+ - \zeta_-}
\right) \, .
\end{eqnarray}
These expressions are identical except for the $\pm$  sign in front of the $\zeta_0^{n+1}$ term. Assuming
that the common link failure rate is $\lambda$ while that for the nodes is $\xi$, the failure frequency for
the $S_0 \rightarrow T_n$ connection is
\begin{equation}
\overline{\nu} = \lambda \, p \, \frac{\partial {\mathcal R}_{T_n}(p,\rho)}{\partial p} + \xi \, \rho \,
\frac{\partial {\mathcal R}_{T_n}(p,\rho)}{\partial \rho} \, ;
\end{equation}
a similar expression applies to ${\mathcal R}_{S_n}(p,\rho)$. When $n$ is large, both availabilities are
actually of the form $\alpha_+ \, \zeta_+^n$, because the modulus of $\zeta_+$ is larger than that of the
remaining eigenvalues for $0 \leq p \leq 1$ \cite{Tanguy06a}. When nodes are perfect, we have therefore in
this limit
\begin{equation}
\overline{\nu} \approx \lambda \, p \, \left( \frac{\partial \alpha_+}{\partial p} \,  \zeta_+^n + n \,
\alpha_+ \, \frac{\partial \zeta_+}{\partial p} \,  \zeta_+^{n-1}  \right) \, ,
\end{equation}
so that the failure rate is
\begin{equation}
\overline{\lambda} \approx \lambda \, \left( \frac{\partial \ln \alpha_+}{\partial \ln p} + n \,
\frac{\partial \ln \zeta_+}{\partial \ln p}  \right) \, , \label{lambdabarre}
\end{equation}
with
\begin{eqnarray}
\frac{\partial \ln \zeta_+}{\partial \ln p} & = & \frac{-1 +4 \, p -6 \, p^2 +4 \, p^3 +(3 - 4 \, p) \,
\sqrt{1+4
\, p^2 \, (1-p)^2}}{2 \, (1-p) \, \sqrt{1+4 \, p^2 \, (1-p)^2}} \, , \label{DLogZeta}\\
\frac{\partial \ln \alpha_+}{\partial \ln p} & = & \frac{4 - 5 \, p +8 \, p^2 -20 \, p^3 +16 \, p^4 -4 \, p^5
-(4 - 7 \, p +4 \, p^2 -2 \, p^3) \, \sqrt{1+4 \, p^2 \, (1-p)^2}}{2 \, (1-p) \, [1+4 \, p^2 \, (1-p)^2]} \,
. \label{DLogAlpha}
\end{eqnarray}
The variations with $p$ of $\partial \ln \zeta_+/\partial \ln p$ and $\partial \ln \alpha_+/\partial \ln p$
are displayed in Figs.~\ref{LambdaBarre3} and \ref{LambdaTilde}. Since $|\zeta_+| <1$ for $p<1$, the
contribution of $\partial \ln \alpha_+/\partial \ln p$ will prevail, and $\overline{\lambda}$ will have a
linear dependence with $n$ in the large network limit (this is a general property when the eigenvalue of
highest modulus is different from unity).

\vskip0.0cm

\vskip0.0cm

For very reliable components, eqs.~(\ref{lambdabarre}--\ref{DLogAlpha}) simplify. It is easy to show by a
series expansion in the link unavailability  $q$ that the global failure rate is given (to first order) by
$\overline{\lambda} \to (2 \, n+4) \, \lambda \, q$; this result could also have been obtained by visual
inspection and enumeration of the minimal cuts.

\section{Conclusion and outlook}

We have shown that the linear differential method for computing the failure frequency is a very simple and
useful one for $k$-out-of-$n$ systems as well as the two-terminal availability for recursive networks (this
should hold for the all-terminal availability, too \cite{Tanguy06c}). Its application is not limited to the
case of extremely reliable components. Even though we restricted our discussion to expressions dealing with
availabilities, a similar treatment could be performed for expressions where unavailabilities are the input
data (see eq.~(\ref{nuGenerale})). For more complex networks, the size of the transfer matrix increases (for
instance, it is a $13 \times 13$ one for the `street $3 \times n$' of \cite{Yeh02}) but the calculations
remain straightforward. Finally, the expressions given for steady-state availabilities can also be used for
time-dependent systems provided that failures and reparations are still statistically independent events,
because the expressions are formally identical (the availabilities of components must be replaced by the
reliabilities).

\appendices
\section{Proof of eq.~(\ref{GKoutofN:G})}
$A_{k,n}$ is given by
\begin{equation}
A_{k,n} = \sum_{l=k}^{n} \, \left( \! \! \!
\begin{array}{c}
n \\ l
\end{array}
\! \! \! \right) \, p^l \, (1-p)^{n-l} \, .
\end{equation}
The fundamental equality between binomials
\begin{equation}
\left( \! \! \!
\begin{array}{c}
n+1 \\ l
\end{array}
\! \! \! \right) = \left( \! \! \!
\begin{array}{c}
n \\ l
\end{array}
\! \! \! \right) + \left( \! \! \!
\begin{array}{c}
n \\ l-1
\end{array}
\! \! \! \right)
 \,
\end{equation}
leads to
\begin{equation}
A_{k,n+1} = (1-p) \, A_{k,n} + p \, A_{k-1,n} \, .
\end{equation}
Setting $\displaystyle {\mathcal G}_k(z) = \sum_{n=0}^{\infty} \, A_{k,n} \, z^n$ implies
\begin{equation}
\frac{1}{z} \, {\mathcal G}_k(z) = (1-p) \, {\mathcal G}_{k}(z) + p
\, {\mathcal G}_{k-1}(z) \, .
\end{equation}
so that
\begin{equation}
{\mathcal G}_k(z) = \frac{p \, z}{1 - (1-p) \, z} \, {\mathcal
G}_{k-1}(z) = \left( \frac{p \, z}{1 - (1-p) \, z} \right)^k \,
{\mathcal G}_{0}(z) \, .
\end{equation}
Since $A_{0,n} = 1, \forall n$, ${\mathcal G}_{0}(z) = 1/(1-z)$; eq.~(\ref{GKoutofN:G}) follows.

%\section*{Acknowledgment}

%CT : Nobody. ADV : ???


\begin{biographynophoto}{Annie Druault-Vicard} joined France Telecom
Division R\&D in 2000. She obtained a PhD in computer science from
Institut National de Recherche en Informatique et Automatique
(INRIA-Rocquencourt) in 1999.
\end{biographynophoto}

\begin{biographynophoto}{Christian Tanguy} joined the Centre
National d'\'{E}tudes des T\'{e}l\'{e}communications in 1987, and
worked in the Laboratoire de Bagneux from 1990 until 2001. He
obtained a PhD in atomic physics from University Paris 6 in 1983 and
a ``Habilitation \`{a} diriger des recherches'' in condensed matter
physics from University Paris 7 in 1996.
\end{biographynophoto}

% insert where needed to balance the two columns on the last page
%\newpage

%\begin{biographynophoto}{Jane Doe}
%\end{biographynophoto}


%\vfill


\end{document}